%% file: main.tex
\documentclass[conference]{IEEEtran}
\IEEEoverridecommandlockouts
\include{user}

\begin{document}
% Authors & Title
\include{authors}

% Title
\maketitle

%%%%%%%%%%%%%%%%%%%%%%%%%%%%%%%%
%
% Abstract
%
%%%%%%%%%%%%%%%%%%%%%%%%%%%%%%%%
\begin{abstract}
The progressive displacing of conventional generation in favor of renewable energy sources requires restoring an adequate capacity of regulating power to ensure reliable operation of power systems. \emph{Battery Energy Storage Systems} (BESSs) are considered to be promising assets to restore suitable frequency regulation capacity levels. BESSs are typically connected to the grid with power-converters, able to operate in either grid-forming or grid-following modes. This paper quantitatively assesses the impact on the local distribution grid of BESSs providing frequency regulation to bulk power systems. Specific metrics are proposed to compare the performance of grid-forming and grid-following control. Experimental results are obtained taking advantage of a 720 kVA/500 kWh BESS connected to the 20 kV distribution grid of the EPFL campus. The quantitative evaluation based on suitably proposed metrics confirms the superior performance of grid-forming strategy, compared to grid-following one. %This is consistent with what previously proved in the literature for low-inertia power grids via simulations.

\end{abstract}
% First page
\IEEEpeerreviewmaketitle

%%%%%%%%%%%%%%%%%%%%%%%%%%%%%%%%
%
% Introduction
%
%%%%%%%%%%%%%%%%%%%%%%%%%%%%%%%%
\section{Introduction}
\label{Chapter:Introduction}
Integrating renewable energy sources in the European energy supply mix has long been ascertained as a fundamental step for the decarbonisation of the electric sector. Moving towards this direction raises multiple challenges for power systems, including the need for regulating power and flexibility services provision to mitigate the massive integration of non-dispatchable and stochastic resources. In this respect, converter-interfaced \emph{Battery Energy Storage Systems} (BESSs) are considered as  perspective assests for grid frequency regulation %(e.g., \cite{zecchino_optimal_2021}) 
in view of their large ramping rates, high round-trip efficiency and commercial availability \cite{dunn_electrical_2011,sossan_achieving_2016,mercier_optimizing_2009}. BESSs interface with the grid through four-quadrant power converters \cite{wang_review_2016}, typically controlled to provide grid ancillary services ranging from frequency regulation up to energy management \cite{namor_control_2019}. As known, there are two main approaches to control converter-interfaced BESSs: \emph{Grid-Forming} (GFR) and \emph{Grid-Following} (GFL).
Nowadays, the majority of converter-interfaced resources are controlled as GFL sources, as this operating mode is considered to be sufficiently efficient~\cite{beerten_modeling_2014,baroudi_review_2007}.
Nonetheless, future grids may require GFR devices to provide more frequency and voltage regulation, stability and black-start capabilities, all services
that are nowadays offered primarily by synchronous machines~\cite{paolone_fundamentals_2020,milano_foundations_2018}. Furthermore, recent studies have proved GFR control strategies to outperform GFL in terms of frequency regulation performance in low-inertia power grids \cite{zuo_performance_2021}. The impact of GFR converters on the dynamics of a reduced-inertia grid has been investigated in \cite{zuo_effect_2020}, which quantitatively proved the good performance of GFR units in limiting the frequency deviation and in damping the frequency oscillations in case of large contingencies. 
The existing scientific literature lacks of studies on the assessment of the performance of GFR units in supporting the frequency containment process of large interconnected power grids. In fact, studies on the GFR units synchronizing with AC power grids are mostly limited to either simulation~\cite{zuo_effect_2020,zuo_performance_2021,liu_comparison_2016} or to experiments on ideal slack buses with emulated voltage \cite{qoria_pll-free_2020,rosso_robust_2019}.
The latter are capable of identifying the fundamental characteristics of the GFR controls, as well as to optimize optimal control parameters \cite{qoria_pll-free_2020}. However, they can not assess the dynamic interaction between power grids and the GFR units. Simulations-based studies have been investigating the impact of large-scale GFR or GFL units on frequency containment of low-inertia power grids\cite{zuo_performance_2021,liu_comparison_2016}, where the power rating of the controlled element (i.e. the BESS) is comparable to the one of the power system. Nevertheless, in bulk power grids, the influence of a single frequency control provider can hardly be detected through frequency measures at the transmission level. An alternative approach is to measure the influence of units providing frequency regulation services on the local distribution grid.
In this respect, the scope of this paper is to experimentally assess the performance of GFR and GFL control for converter-interfaced BESS by measuring with accurate \emph{Phase Measurement Units} (PMUs) the influence of the two control strategies on the local distribution grid.
% and to propose an experimental validation. %In particular, the study proposes a new metric, able to measure the local effects of the two control strategies at the feeder level. Moreover, an experimental comparison of GFR and GFL control local performance is proposed, and the observed effect is compared with the one presented in the literature. 
The experimental campaign is conducted by taking advantage of a grid-connected 720 kVA/500 kWh BESS connected to the 21 kV distribution feeder on the EPFL campus.%, and the results are compared with the one presented in the literature. 

The paper is structured as follows: \cref{Chapter:SoA} presents the characteristics GFR and GFL, \cref{Chapter:KPI} analyses the state-of-the-art metrics used to compare GFR and GFL control and proposes a new metric, to better capture local effects of the two control strategies. \cref{Chapter:Result} presents and discusses the experimental results. Finally, \cref{Chapter:Conclusion} summarizes the results and concludes the paper.

%%%%%%%%%%%%%%%%%%%%%%%%%%%%%%%%
%
% Section 2: State of the art
%
%%%%%%%%%%%%%%%%%%%%%%%%%%%%%%%%

\section{Converter Controls}
\label{Chapter:SoA}
As mentioned in \cref{Chapter:Introduction}, GFR and GFL controls are the two approaches currently used to control converter-interfaced BESSs~\cite{paolone_fundamentals_2020,rocabert_control_2012}. 
For the sake of consistency, this section reports the description of GFR and GFL controller proposed by \cite{paolone_fundamentals_2020}.
%Here below we recall their definitions proposed in~\cite{paolone_fundamentals_2020}. 

\begin{figure}[t]
	\centering
	\subfloat[Main scheme of a GFR control.]{\includegraphics[width=0.8\columnwidth]{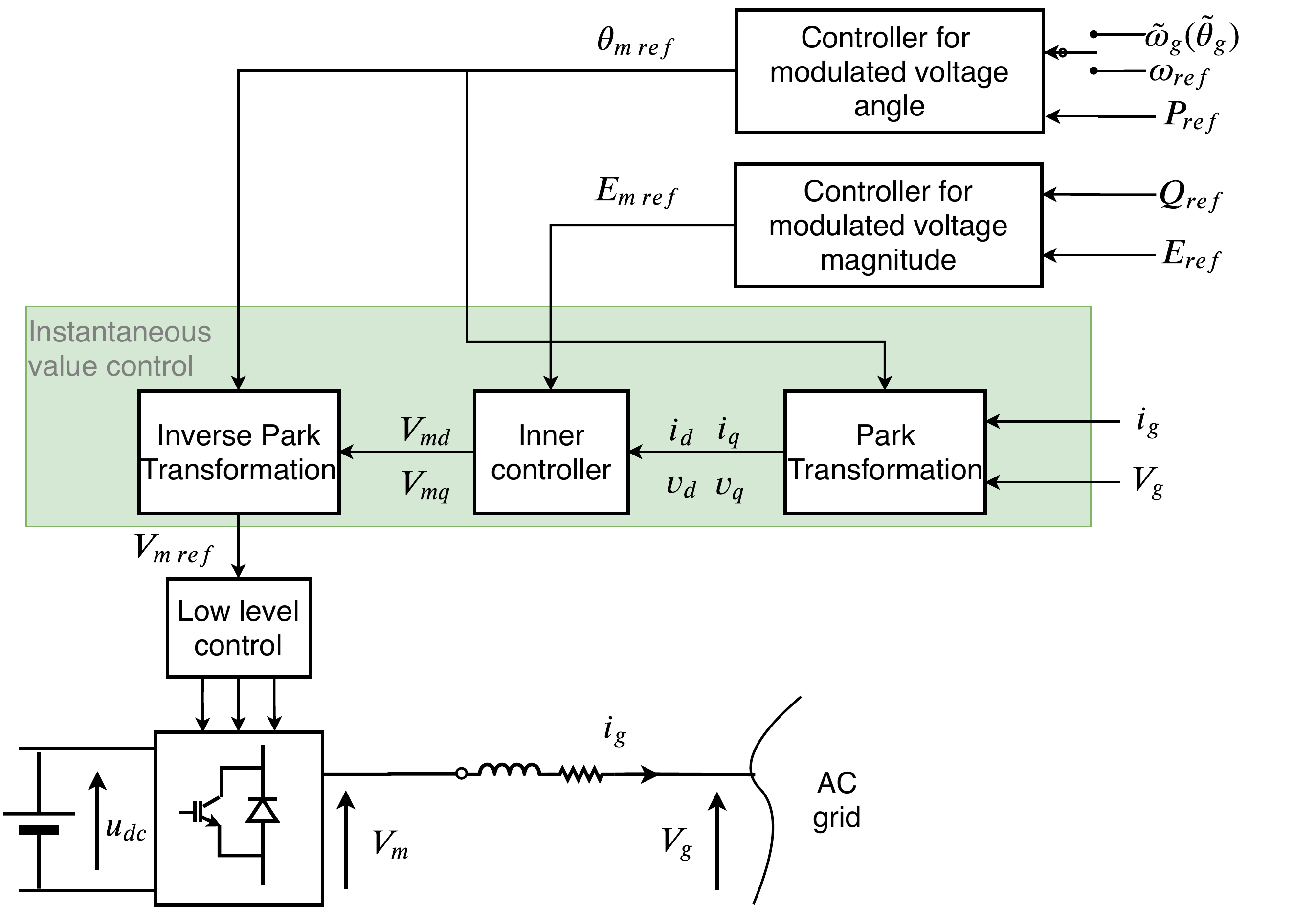}\label{fig:Schem_GFM}}
	\quad
	\subfloat[Main scheme of a GFL control.]{\includegraphics[width=0.8\columnwidth]{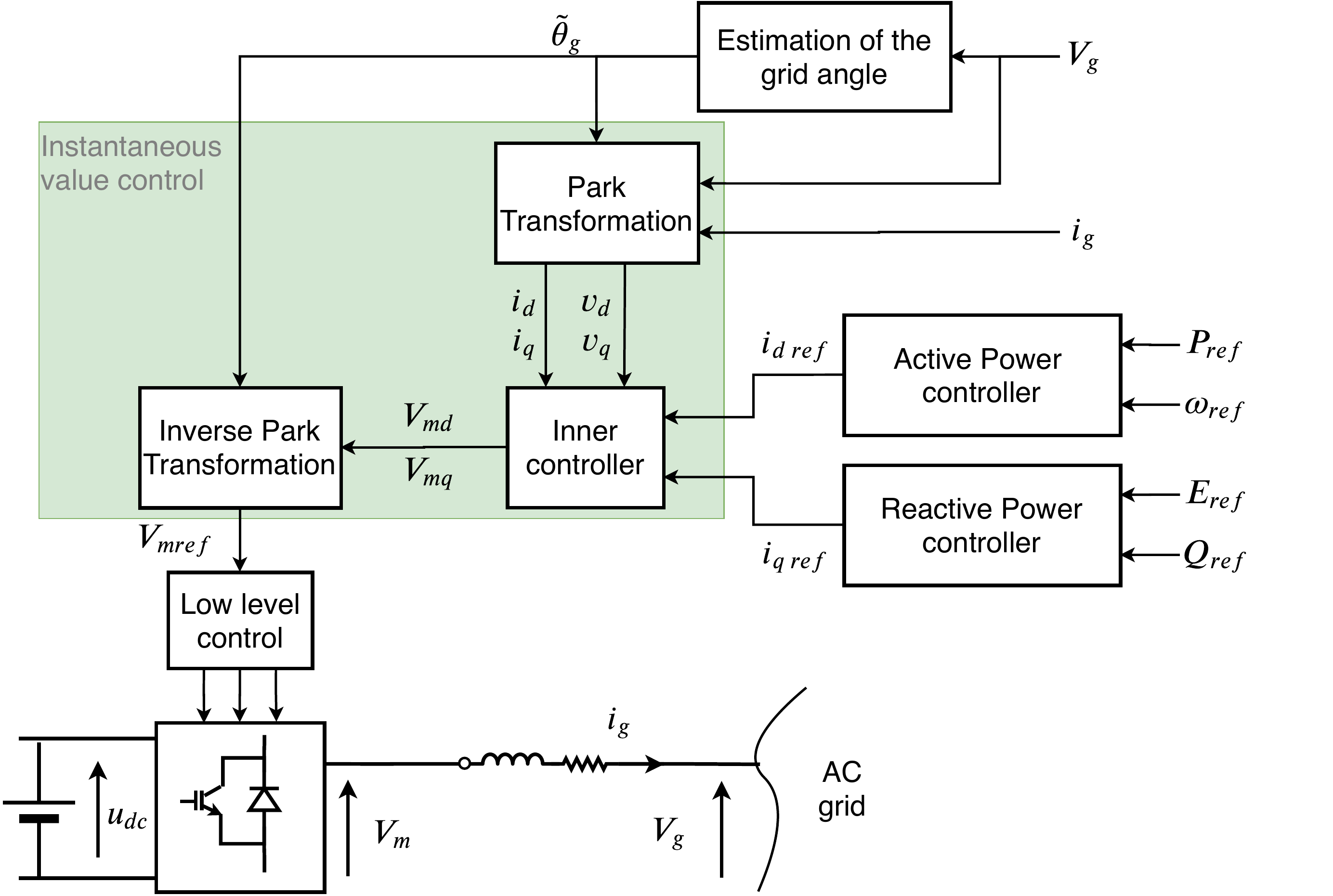}\label{fig:Schem_GFL}}
	\caption{Main schemes of GFR and GFL controls %The diagrams are inspired from~
	\cite{paolone_fundamentals_2020}.}
	\label{fig:schemGFMGFL}
\end{figure}

GFR units are based on a power converter which controls magnitude and angle of the voltage at the \emph{Point of Battery Coupling} (PBC). As a consequence, the knowledge of the fundamental frequency phasor of the grid voltage at the PBC is not strictly necessary. 
%Depending on the characteristics of the network to which the converter is connected, an isolated system or a slack bus, it is possible by means of outer control loops to adapt the injected instantaneous active power as well as to provide voltage and frequency support.
Fig.~\ref{fig:Schem_GFM} presents a general structure of the GFR control, where the modulated voltage angle is linked with the converter active power, while the voltage magnitude is regulated by taking into account of voltage or/and reactive power reference. 

GFL units are based on a power converter which controls the currents injection to adjust the active and reactive power injection. To this end, the injected current is controlled with a specific phase displacement with respect to the grid voltage at the PBC. Consequently, the fundamental frequency phasor of the grid voltage at the PBC is needed at any time to correctly calculate the converter reference currents. 
%The amplitude and angle of the reference currents with respect to the grid voltage phasor are properly modified by outer control loops so as to inject the required amount of active and reactive power. 
\cref{fig:Schem_GFL} shows the classical structure of the GFL control, where the grid voltage angle $\tilde{\theta}_{g}$ is estimated thanks to a \emph{Phase Locked-Loop} (PLL) and used by the Park's transformation.
The outer control loops can adjust the amplitude and angle of the current reference currents to inject required amount of active and reactive power.  
Grid-supporting GFL converters can be applied to adjust active and reactive power set points under frequency or voltage deviations

%The active current reference, $I_{d~ref}$, is generated by an outer active power loop, where the frequency can participate to enable frequency supporting (e.g. frequency containment support). The reactive current reference, $I_{q~ref}$ is the output of a reactive power loop, where voltage can participate to enable voltage supporting. With respect to the active and reactive current references (i.e., $I_{d~ref}$ and $I_{q~ref}$), the inner controller respectively generates the d- and q-axis components of the modulated voltage to be achieved by the converter. 

%In~\cite{zuo_effect_2020}

%%%%%%%%%%%%%%%%%%%%%%%%%%%%%%%%
%
% Section 3: Metrics
%
%%%%%%%%%%%%%%%%%%%%%%%%%%%%%%%%

\section{Quantification of Frequency Regulation}
\label{Chapter:KPI}
%In order to quantitatively evaluate the performance of GFR units proving frequency regulation, different metrics can be used. In the first part of this section, metrics previously defined in \cite{zuo_performance_2021} %for the assessment of frequency regulation performance in low-inertia grids 
%are discussed. %Their application to bulk power system is discussed. 
%As most of them are unable to capture the effects of frequency control action from a single element of the grid, new evaluation metrics are proposed.  %The metrics can be used to quantitatively compare the GFR vs the GFL control, as well as be generalized to different size of converter-interfaced units. 
\subsection{State-of-the-Art Metrics}
To evaluate the impact of the frequency regulation provided by BESSs to low-inertia grids, ~\cite{zuo_performance_2021} proposes the hereby listed metrics:
\begin{itemize}
    \item Frequency \emph{Probability Density Function} (PDF), where the standard deviation of the grid-frequency statistical distribution is observed to assess the capability of the control strategy in limiting the frequency deviations from the nominal value.
    \item \emph{Integral Frequency Deviation} (IFD):
    \begin{align}
        \text{IFD} = \sum_i^M\sum_{k=1}^N{|f_{k,i}-f_0|}
    \end{align}
    where $M$ is the number of measurement units, $N$ is the total sampling number of frequency measurements for each load, $f_{k,i}$ is the frequency measured by the measurement unit $i$ at sample $k$ and $f_0$ is the nominal frequency, i.e. 50 Hz.
 \end{itemize}
Both frequency PDF and IFD establish a relation between grid-frequency measurements %(with a reporting rate of 50 frames per second) 
%\cite{romano_integration_2015} 
and control actions of the converter-interfaced BESS. Nevertheless, it is not possible to measure the influence of a  small frequency control provider to the bulk power grid, as the frequency can be considered independent from the single action of a small power converter. Therefore, frequency PDF and IFD can not be considered as reliable metrics.

\subsection{New metrics}
To characterize the local effect of a converter-interfaced BESS providing frequency regulation and to identify the difference between the GFR and GRL controls, we proposed new metrics based on PMUs measurements performed locally, i.e., on the distribution feeder where the BESS is installed.
The frequency measures at the \emph{Point of Common Coupling} (PCC) of the distribution feeder, as well as the phase angle differences between the voltage controlled by the converter unit and the PCC voltage % of the upper-level grid 
are taken into consideration to derive the following metrics:
\begin{itemize}
    \item {\emph{Relative Rate-of-Change-of-Frequency} (rRoCoF):}
\begin{align}
rRoCoF = \left | \frac{\Delta f_{PCC} / \Delta t}{\Delta P}\right |
\end{align}
where  $\Delta f_{PCC}$ is the difference between one grid frequency sample and the next (once-differentiated value) at the bus where the BESS is connected to, $\Delta P$ is the once-differentiated BESS active power, and $\Delta t$ is the sampling interval. 
\item {\emph{Relative Phase Angle Difference Deviation} (rPADD):}
\begin{equation}
    rPADD_k = \left | \frac{\Delta \theta_k -  \Delta \theta_0}{\Delta P_k}\right |
    \label{eq:PADD}
\end{equation}
%\begin{equation}
%\begin{cases}
%\Delta \theta_k = \Delta \theta_{k,PMU1} - \Delta \theta_{k,PMU2} \\
%\Delta \theta_0 = \Delta \theta_{0,PMU1} - \Delta \theta_{0,PMU2} \\
%\end{cases}
%\end{equation}
quantifying the change in the phase-to-neutral voltage angle difference $\Delta \theta_k$, measured by two PMUs installed on nodes at different voltage level of the local feeder, versus the case with null delivered active power $\Delta \theta_0$.
\end{itemize}
The relative RoCoF metric measures the RoCoF regulation at PCC. Since the RoCoF is weighted by the delivered active power of the BESS, it can be used to compare the local performance of GFR versus GFL converter in large interconnected electrical systems.
On the other side, for the rPADD computation, the benefit of using metrics based on voltage phase angle is two-fold:
\begin{enumerate*}[label=(\roman*)]
\item changes in the phase angle estimation are directly linked with frequency deviations from nominal frequency \cite{noauthor_ieee_2011} and
\item it can indicate the ability of the connected unit in sustaining local voltage by controlling its angle.
 \end{enumerate*} The metric is normalized by the delivered active power of the converter-interfaced unit to ensure its independence from the grid condition and the power rating of the converter-interfaced unit.

%%%%%%%%%%%%%%%%%%%%%%%%%%%%%%%%
%
% Section 4: Experiments
%
%%%%%%%%%%%%%%%%%%%%%%%%%%%%%%%%

\section{Experimental campaign}
\label{Chapter:Result}
\subsection{Experimental Setup}
\subsubsection{Feeder Characteristics}
\begin{figure}[t]
    \centering
    \includegraphics[width=0.7\linewidth]{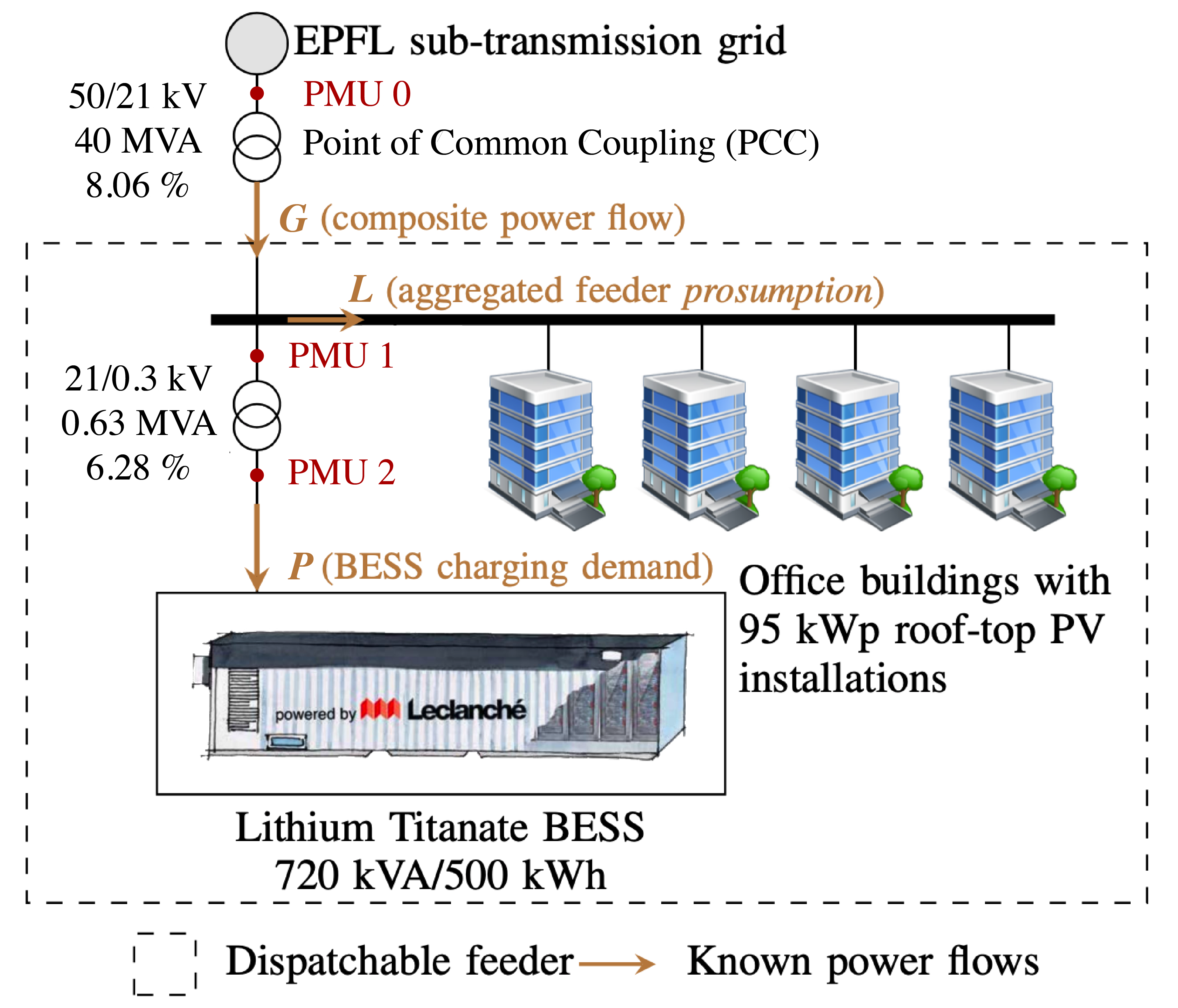}
    \caption{Experimental setup of the EPFL campus MV grid hosting the utility-scale 720kVA/500 kWh BESS.}
    \label{fig:ExperimentalSetup}
\end{figure}

% Feeder description 
for the experimental campaign, a 20 kV distribution network equipped with a BESS is considered. The experimental configuration shown in \cref{fig:ExperimentalSetup} consists in a group of buildings of the EPFL campus characterised by a  140 kW base load, hosting 95 kWp root-top PV installation and a grid-connected 720 kVA/500 kWh Lithium Titanate BESS. The BESS bidirectional real power flow is denoted by $P$, and $G$ is the composite power flow seen at the PCC. 
The aggregated building demand is denoted by $L$ and, by neglecting grid losses \cite{pignati_real-time_2015} %\footnote{The targeted grid has a radial topology and characterized by co-axial cables line with a cross section of 95 mm$^2$ and a length of few hundreds meters. Therefore, the grid losses are negligible.  See \cite{pignati_real-time_2015} for further details.}
, it is estimated as $L=P-G$. For both buildings and BESS, positive values denote injection towards the PCC, while negative denote consumption. The measuring systems is composed by  three PMUs, i.e. PMU 2, PMU 1 and PMU 0, installed respectively on 0.3 kV, 20 kV and 50 kV side of the feeder.  Two transformers connect the different voltage levels of the feeder. Their nominal voltage and power, as well as their short circuit voltage expressed in percent are shown in \cref{fig:ExperimentalSetup}.
% BESS description 
\subsubsection{Controller Characteristics}
the BESS converter is capable of synchronizing with the grid in both GFL and GFR control modes. In GFR mode the converter controls the voltage angle to provide frequency regulation by means of an internal control loop directly establishing a droop relation between the grid frequency and the active power to be delivered. In GFL mode, the active power set-point for frequency regulation is computed in an external loop, as a function of the grid frequency deviation (from 50 Hz). In particular, the maximum feasible droop, i.e. 1.44 MW/Hz, is selected for the experiments and implemented in both GFR and GFL modes to maximise the BESS response.

% PMU description
\subsubsection{Requirement for PMUs}
for the computation of the Relative Integrated Phase Angle Difference Deviation (rPADD), high-accuracy PMUs are required. Knowing the feeder topology, the short-circuit voltage of the two transformers and assuming the line as lossless, it is possible to estimate the angle displacement per unit of delivered power by the BESS as 0.006 degree/kW. Considering a droop of 1.44 MW/Hz and the need of estimating BESS reactions for frequency deviations greater than the typical frequency deadband \cite{noauthor_commission_nodate}, i.e. 10 mHz \cite{noauthor_commission_nodate}, the measuring infrastructure must be able to detect angle differences of 0.9 degree. %To robustly measure those phase angle differences, the latter have to be one order of magnitude larger than the PMU accuracy confidence interval $3\sigma$. 
To robustly measure those phase angle differences, the PMU-based distributed sensing with infrastructure deployed in the EPFL campus have been utilized \cite{pignati_real-time_2015}, thanks to its accuracy in terms of 1 standard deviation equal to $0.001$~degrees (i.e. 18~$\mu$rad), 90 times smaller than the minimum angle required for the rPADD computation.
%As shown in \cref{fig:ExperimentalSetup}, 2 PMUs (PMU0 and PMU1) at the different voltage levels of 50 kV and 21 kV were utilized to compute the difference in the voltage phase angles at the two sides of the impedance in-between (line + 50kV/21kV transformer) when a given power was provided by the BESS. 
%%%%%%%%%%%%%%%%%%%%%%% Ask tony about this part %%%%%%%%%%%%%%%%%%%%%%%%%
%A first set of experimental tests showed that when operating the BESS with the maximum possible droop of 1.44 MW/Hz, the BESS active power was not large enough to guarantee meaningful dis- placements of the phasors at the two measurements points, given the PMU capabilities in terms of accuracy and precision described in Section 2. For this reason, it was decided to consider a larger impedance between the two measurement points, which could provide larger displacements of the phasors for the same amount of BESS active power. In particular, a new PMU (PMU2) was installed at the 300 V LV side of the BESS step-up transformer. This enabled the comparison of the 2 measurements with a higher impedance in-between, given the high value of the impedance of the BESS step-up transformer, as detailed in \cref{fig:ExperimentalSetup}. 

\subsubsection{Grid condition}
as the experimental validation relies on real operating conditions, it is subjected to the uncontrollable variation of the power system frequency of the Continental Europe Synchronous Region, to which the medium-voltage grid of the EPFL campus is connected. For this reason, the experimental campaign has taken place during the transition of the hour, which typically present fast frequency ramps due to mechanisms related to electricity markets \cite{weibbach_impact_2018}. Such frequency ramps have been used to activate BESS to perform considerable active power exchanges $P$, following the 1.44 MW/Hz droop implemented both in the GFR and  GFL controls. Furthermore, the proposed tests have been carried out when the base consumption of the local medium voltage $L$ grid presents a mostly constant profile such that its dynamics are not influencing the local measures on which the metric are based, i.e., during evening and night hours. This allows for assuming $\Delta\theta_0$ to be constant during the experiment, enabling a better identification of the impact of the BESS provided power in the local phase angle displacements.

\subsection{Experimental Results}

\begin{figure}[t]
    \centering
    \includegraphics[width=0.9\linewidth]{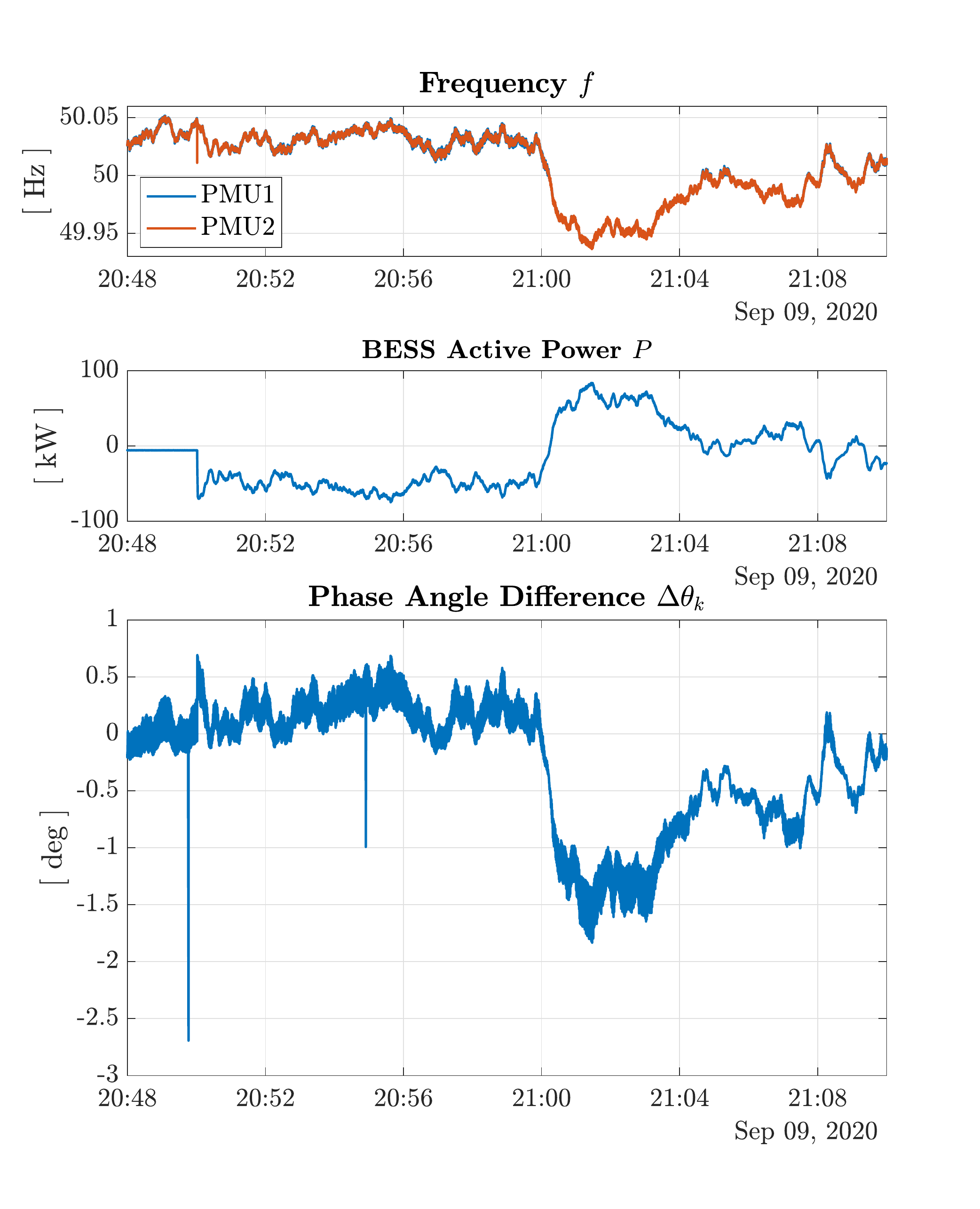}
    \vspace{-0.5cm}
    \caption{Experimental test results using the EPFL BESS converter in GFR mode with 1.44 MW/Hz droop.}
    \label{fig:GFR}
\end{figure}

\begin{figure}[t]
    \centering
    \includegraphics[width=0.9\linewidth]{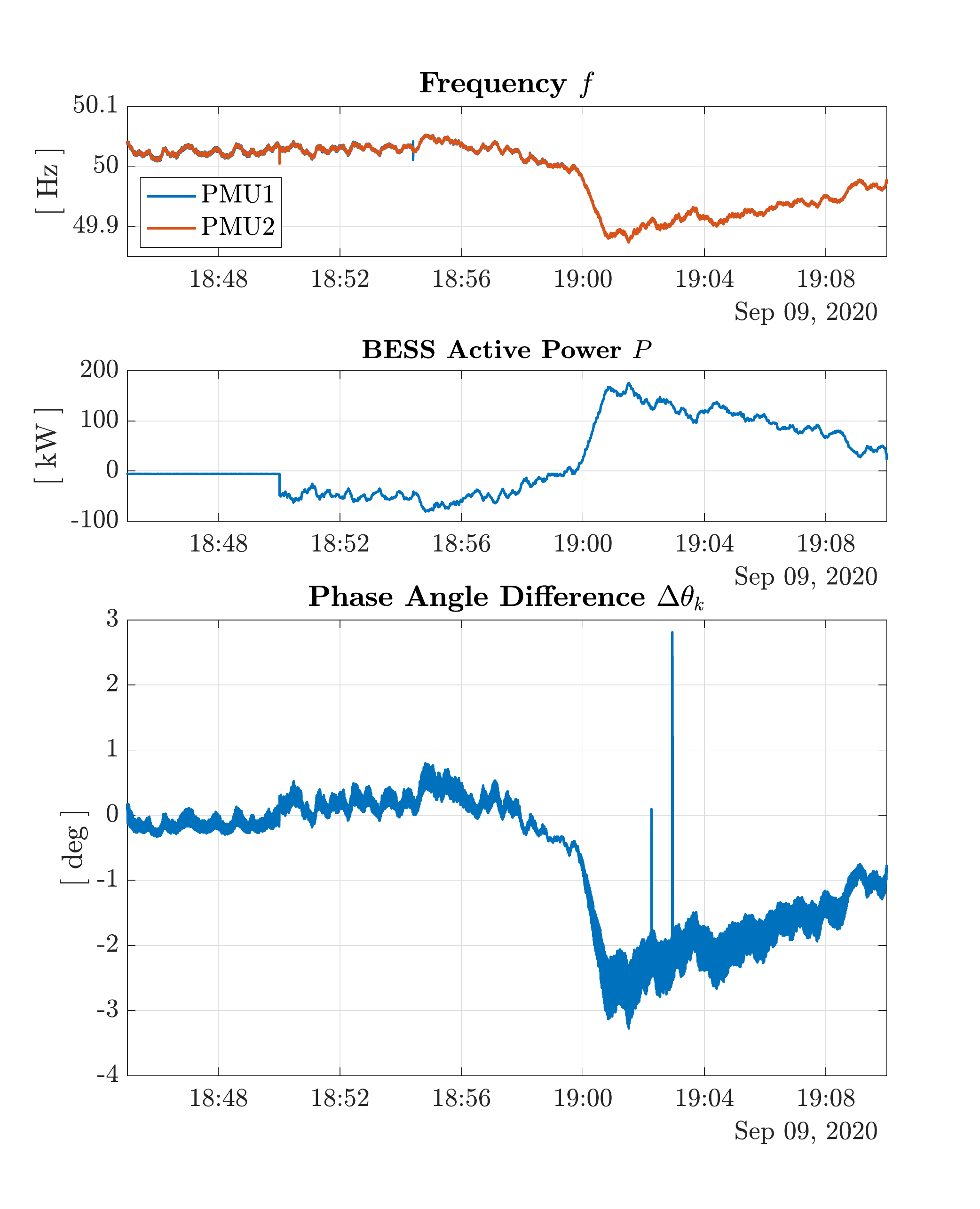}
    \vspace{-0.5cm}
    \caption{Experimental test results using the EPFL BESS converter in GFL mode with 1.44 MW/Hz droop}
    \label{fig:GFO}
\end{figure}

The experimental assessment of the BESS response are shown in \cref{fig:GFR}. At the top, the frequency measured by the PMUs and the BESS active power response are provided. The frequency ramp at the hour transition (21:00 CET) are noticeably reflecting on the BESS power ramp. 
The angle displacement is computed as the angle difference between PMU~1 (21 kV) and PMU~2 (0.3 kV): 
\begin{equation}
    \Delta \theta_k=\theta_{k, PMU1}-\theta_{k, PMU2}
\end{equation}

Before the activation of the BESS control at 20:50 CET, the phase angle difference $\Delta \theta_0$ is computed as:
\begin{equation}
\Delta \theta_0=\theta_{0, PMU1}-\theta_{0, PMU2}
\end{equation}
The values of $\Delta \theta_0$ and $\Delta \theta_k$ can be inserted in \cref{eq:PADD} for the rPADD computation. For the sake of comparison, a similar experimental assessment is conducted in GFL mode and the results are shown in \cref{fig:GFO}.

%\begin{figure}[t]
%    \centering
%    \includegraphics[width=0.8\linewidth]{Fig/GFR.png}
%    \caption{Experimental test results using the EPFL BESS converter in GFR mode with 1.44 MW/Hz droop.}
%    \label{fig:GFR}
%\end{figure}
%\begin{figure}[t]
%    \centering
%    \includegraphics[width=0.8\linewidth]{Fig/GFO.png}
%    \caption{Experimental test results using the EPFL BESS converter in GFL mode with 1.44 MW/Hz droop}
%    \label{fig:GFO}
%\end{figure}

%On the left side, the frequency measured by the PMUs and the BESS active power response are provided. At the hour transition (19:00) a frequency ramp can be noticed, as well as the BESS  active power ramp provision. On the right side, the angle displacement is computed as the angle difference between PMU~1 (21 kV) and PMU~2 (0.3 kV): 
% according to the definition included in \cref{Chapter:KPI}. It is observed that the angle displacement $\Delta \theta_k$ is stable before 18:50, i.e., before the BESS starts being controlled, allowing for the computation of:
%Similarly, Fig.~\ref{fig:GFR} shows the results of the experimental assessment of the BESS response when operated in GFR mode with the maximum $f-B$ droop (1.44 MW/Hz) allowed by the available converter technology. 

\begin{figure}[t]
    \centering
    \includegraphics[width=0.95\linewidth]{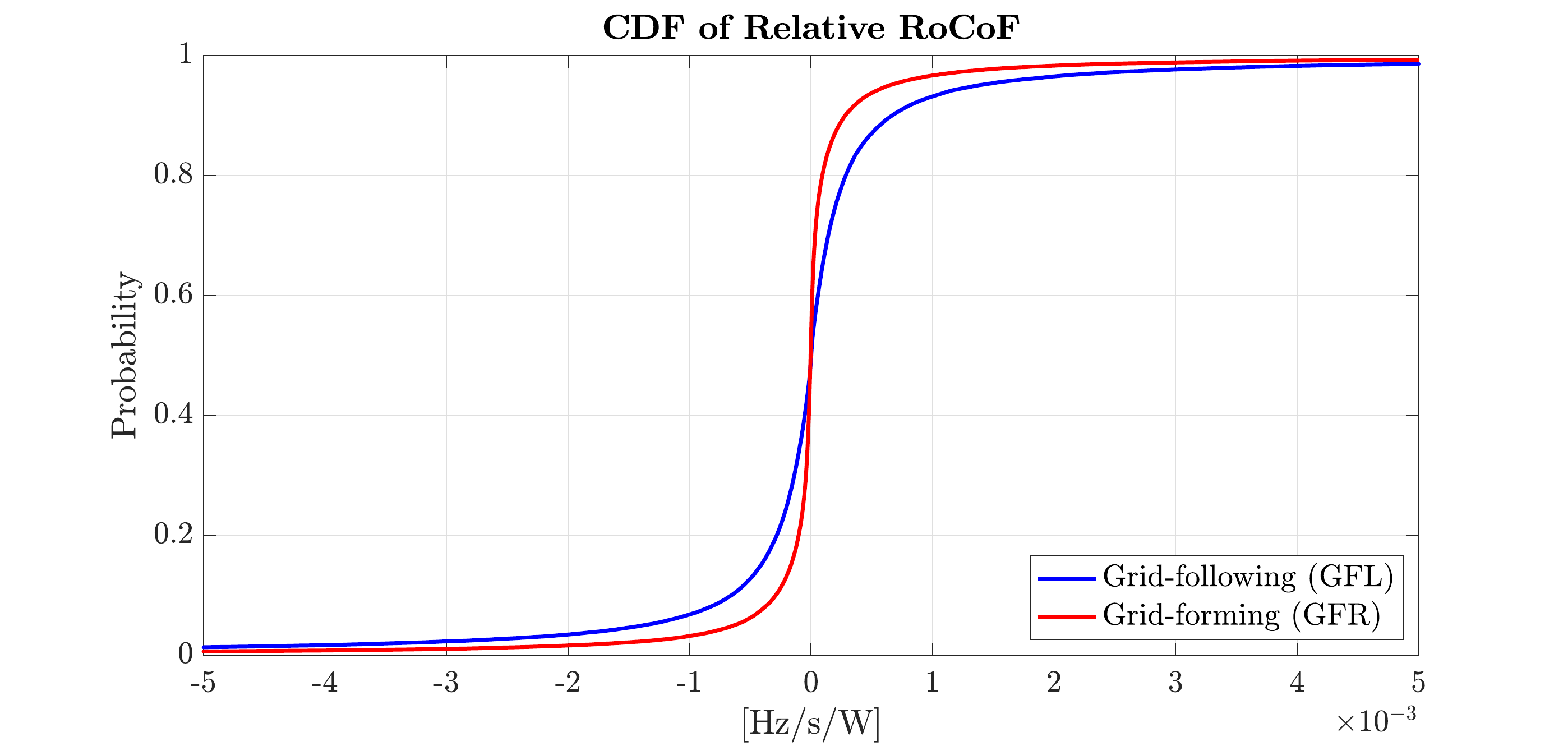}
    \caption{Cumulative distribution function of the rRoCoF [Hz/s/W] for the experimental tests.}
    \label{fig:rRoCoF}
\end{figure}

\begin{figure}[t]
    \centering
    \includegraphics[width=0.95\linewidth]{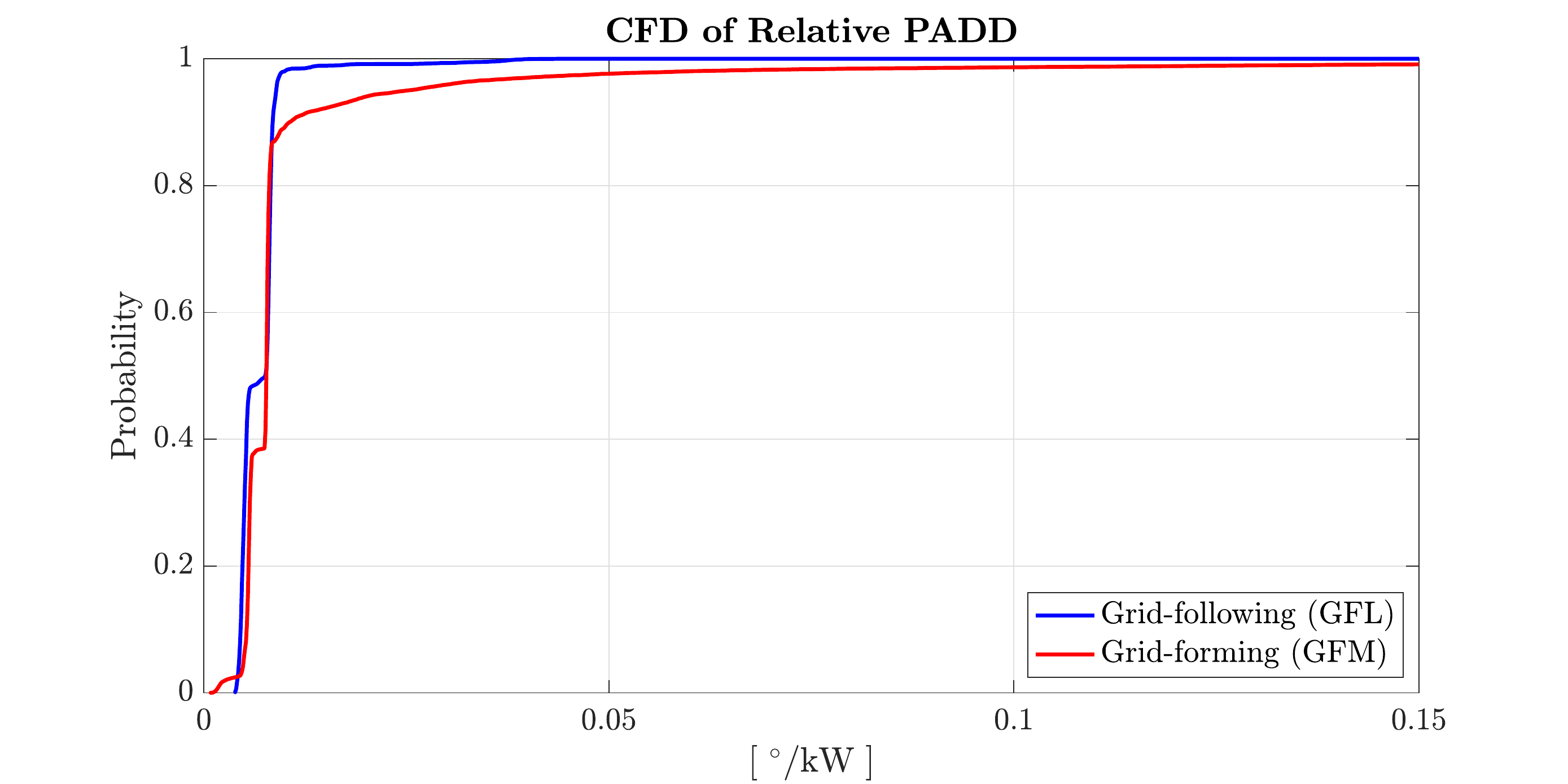}
    \caption{Cumulative distribution function of the rPADD [deg/kW] for the experimental tests.}
    \label{fig:PADD}
\end{figure}

To quantitatively assess and compare the two cases with respect to the impact on the local grid frequency, the two relative metrics presented in \cref{Chapter:KPI}, namely rRoCoF and rPADD, are considered. rRoCoF is computed over a time window of 60 ms. For the sake of clarity and simplicity of the interpretation, results of the rRoCoF the rPADD computation are reported in terms of \emph{Cumulative Distribution Function} (CDF).
Fig. \ref{fig:rRoCoF} shows the rRoCoF CFD for multiple tests, expressed in [Hz/s/kW]. It is important to observe the vertical trend of the red line, representing the GFR case, implying higher probabilities of having smaller values of rRoCoF. 
This means that, for a higher number of instances for which the rRoCoF has been computed, it is more likely to have smaller values when controlling the BESS in GFR mode than in GFL mode. This demonstrates that the GFR control can guarantee larger containment of the rate of change of the frequency, compared to the more traditional GFL strategy. These results are in line with what is presented in \cite{zuo_performance_2021}, relatively to the simulation activities on the IEEE 39-bus benchmark grid.
Fig. \ref{fig:PADD} shows the CDF probability of the rPADD of having a given value expressed in [deg/kW]. It is clear that larger values of the index mean larger impact on the local phase angle displacement of the voltages measured at the two different voltage levels. It can be noticed that in this case larger values are found for the GFR case, demonstrating that the implemented GFR control can guarantee larger impact on the local phase angle difference, compared to the more traditional GFL strategy. %These results are in line with what was found previously in occasion of the simulation activities on the IEEE 39-bus benchmark grid.


%%%%%%%%%%%%%%%%%%%%%%%%%%%%%%%%
%
% Section 5: Conclusion
%
%%%%%%%%%%%%%%%%%%%%%%%%%%%%%%%%
\section{Conclusion}
\label{Chapter:Conclusion}
A novel metric to quantitatively assess the local impact of BESSs providing frequency regulation to bulk power grid has been proposed.
Experimental tests confirmed the outcome of the analysis carried out by means of simulation activities on the low-inertia configuration of the IEEE 39-bus emulated grid presented in \cite{zuo_performance_2021}, validating the positive effects of the grid-forming strategy in the control of the local frequency. The upgrade of the PMU-based sensing infrastructure, enabled the assessment of impacts on the local frequency with the BESS control running with maximum possible frequency droop. The proposed metrics (rROCOF and rPADD) can potentially be of use for the quantification of frequency reserve provision to distribution systems. Moreover, both metrics confirmed the superior performance of the grid-forming strategy, characterised by a more contained rROCOF and a larger impact in the local phase angle compared to grid-following mode.%when operating in grid-forming mode rather than in grid-following mode, and at the same time larger impacts in the local phase angle were also caused.

%\section*{Acknowledgment}

%\def\url#1{}

%The authors would like to thank... \cite{atta_phasor_2014}
\bibliographystyle{IEEEtran}
\bibliography{IEEEabrv,reference}

\end{document}

%% file: user.tex
\usepackage{amsmath,amssymb,amsfonts}
\usepackage{algorithm}
\usepackage{algorithmic}
\usepackage{graphicx}
\usepackage{textcomp}
\usepackage{xcolor}

\usepackage[caption=false,font=footnotesize]{subfig}	%	The option 'caption=false' is required to keep IEEE-style captions.
\usepackage{graphicx}
\usepackage{amsmath,amsfonts,amssymb,amsthm}
\usepackage{mathtools}
\usepackage{mathrsfs}	%	mathscr
\usepackage{subdepth}	%	uniform subscript position
\usepackage{array}	%	better appearance for array and tabular environments
\usepackage[inline]{enumitem} % enumerate stuff in line
\usepackage{hyperref}
\usepackage[sort,compress,capitalise]{cleveref}	%	...
\usepackage{url}
\usepackage{xcolor} % Coloured text
\usepackage{xspace} % Intelligent spacing (acronym commands) 
\usepackage{lipsum}
\usepackage{tikz}
\usetikzlibrary{babel}  % Avoid problems due to babel library
\usepackage[straightvoltages,nooldvoltagedirection]{circuitikz}
\usetikzlibrary{shapes,arrows}
\usepackage{verbatim}
\usepackage{multirow}
\usepackage[utf8]{inputenc}
\usepackage{fourier} 
\usepackage{array}
\usepackage{makecell}

%% file: authors.tex
%%%%%%%%%%%%%%%%%%%%%%
%%%%%%%%%%%%%%%%%%%%%%
% Title of the Paper
%\title{Effect of Grid-Forming Converters-Interfaced Battery Storage Systems Providing Frequency Regulation in Bulk Power Grids}
\title{Local Effects of Grid-Forming Converters Providing Frequency Regulation \\ to Bulk Power Grids}

% Author names and affiliations
\author{
%\IEEEauthorblockN{Antonio Zecchino}
%\IEEEauthorblockA{SwissGrid AG \\
%Ancillary Services \& Analytics \\
%Aarau, Aargau, Switzerland\\
%antonio.zecchino@epfl.ch
%}
%\and
\IEEEauthorblockN{Antonio Zecchino, Francesco Gerini, Yihui Zuo, \\Rachid Cherkaoui, Mario Paolone}
\IEEEauthorblockA{Distributed Electrical Systems Laboratory (DESL)\\
Ecole Polytechnique Fédérale de Lausanne, EPFL\\
Lausanne, Switzerland\\
\{antonio.zecchino, francesco.gerini, yihui.zuo,\\ rachid.cherkaoui, mario.paolone\}@epfl.ch}
\and
\IEEEauthorblockN{Elena Vagnoni}
\IEEEauthorblockA{Technology Platform for Hydraulic Machines \\ 
Ecole Polytechnique Fédérale de Lausanne, EPFL\\
Lausanne, Switzerland\\
\{elena.vagnoni\}@epfl.ch}
\thanks{{This work is supported by OSMOSE project.  The project has received funding from the European Union's Horizon 2020 Research and Innovation Program under Grant 773406.}}

}